# Self-Checks In Spreadsheets: A Survey Of Current Practice

David Colver,
Director, Operis Group plc
dcolver@operis.com

**ABSTRACT**

*A common application of spreadsheets is the development of models that deliver projections of the future financial statements of companies established to pursue ventures that are subject to project financing. A survey of 11 such spreadsheets prepared by a range of organisations shows that the amount of self-testing included in such models ranges between one formula of testing for each three formulae of calculation, down to essentially no self-testing at all.*

## 1. VANTAGE POINT

Operis, the company of which the author is a director [Operis, 2010], is a specialist advisor in project finance. The term project finance means something particular to bankers, but informally it can be taken to concern transactions involving the exploitation of natural resources or the development of national or regional infrastructure. At the centre of our work, and that of any other firm addressing the same market, is the development of financial models used to structure deals as they are negotiated.

It has become conventional for parties to a project financed transaction to insist that the financial model undergoes independent audit [Croll, 2003] before they finally commit themselves to participating. We estimate to have close to a third of the market for auditing these spreadsheets.

It is from this perspective of developing our own financial models, and reviewing models that other people have developed, that we have observed that there are wide variations in the defensiveness with which financial models are constructed. The purpose of this paper is to offer some suggestions for good practice [Grossman, 2002], and to see how widespread their use is.

## 2. OPERIS STANDARD PRACTICE

We believe there is no one right way to build a financial model [Colver, 2004]. Design choices involve trade-offs that will work well in some circumstances but badly in others.

The method that we have developed as standard practice in our own work involves building up financial models using several worksheets in a single workbook to separate the model's inputs, workings and outputs. The exact layout, and the considerations that motivate them, are expressed in that earlier paper.

For simplicity, that explanation omitted one detail. Among the worksheets is one named Audit. On that is laid out a great number of tests that check that the model doesn't contain various elementary errors.

Each test generates a result that is TRUE if it is passed, and FALSE if it is failed. Those tests are AND'ed together to give an overall result. Conditional formatting is then used, linked to that overall result, such that a disagreeable and unmissable red stripe appears along the top of all the worksheets in the model if any of the tests fail.

## 3. TESTS USED

The tests we use fall into fourteen broad groups:

1   *Balance sheet balances*: Not all spreadsheet models make projections of the future financial statements of businesses, but the majority of the spreadsheets that we prepare do, and the most basic test is that the balance sheet balances.

   (This test is valuable only so far as the elements of the balance sheet are calculated independently, and can then be seen to give consistent totals for assets and liabilities. It isn't useful if one relies on this property of the financial statements to leave one quantity calculated as the balancing item. To our surprise, this shortcut is actively taught and encouraged in certain well-known US business schools, which can cause discomfort between US-trained spreadsheet modellers and European spreadsheet auditors.)

2   *Financial statements add up*: We include tests that compare every subtotal on every output sheet with the sum of its elements. Nothing destroys the credibility of a financial model so quickly as a cash flow statement or a profit and loss account that doesn't add up, even if the fault is cosmetic and doesn't affect any part of the rest of the model.

3   *Financial statements have expected sign*s. This is effective at spotting that loans have been over-repaid, or assets over-depreciated; the sign of the relevant balance sheet entry reverses.

4   *Sources match uses*: Any statement of source and uses of funds shows the quantities in balance.

5   I*dentities hold true*: By identities, we mean expected relationships between different numbers in the model. Many are consequences of the rules of accounting. Examples are that capital costs and depreciation can be reconciled over the life of the project; that accumulated cash and profit, from the cash flow and profit and loss statements, match the corresponding entries in the balance sheet. Others are dictated by the layout of the model: that the operating costs shown on the cash flow matches the total shown at the foot of another schedule that derives those costs from their components.

6   *Balance sheet clears out*: Though not appropriate for a company that continues to exist indefinitely, a project that is subject to a finite life (determined by a finite natural resource, such as an oil deposit, or by a contract of defined length) should show a balance sheet that is full of zeroes at the end.

7   The bottom of the *cash cascade gives the same net cash figure as the cash flow*. (This is a rather technical point, and has consequences for the way the cash cascade is coded, but is a particularly powerful test in practice.)

8   *Ratio inclusion analysis*: We demonstrated inclusion analysis, a technique that increases the chance of detecting acts of omission in ratios, at last year's Eusprig [Colver, 2007], We provide such an analysis for every relevant ratio.

9   *Tax reconciliation*: A test, based on inclusion analysis, that proves the model's tax calculation.

10  *Yield analysis*: Every interest bearing instrument modelled is subject to a simple test to check that the interest charged implicit in the aggregate cash flows can be reconciled to the specified interest rate assumption.

11  *Physical identities*: The calculation of revenues and costs reflects any usable properties of the underlying physical reality. For example, if a chemical plant uses as the feedstock for one part of the facility a chemical that is the output of another section, the two quantities should be tested for equality.

12. *Complete solution*: Any macros used to iterate to a solution have converged.

    (Again there are tensions here between US and European practice: project finance specialists in Europe commonly use macro-controlled iteration to solve financial models because they are taught to avoid circular spreadsheets at any price; some US business schools actively teach the use of circularity.).

13. *Inputs make sense*. Examples are that dates fall within the model's timeline; that enough financing is provided to meet the costs of completing the project.

14. *Outputs meet participants' requirements*: Loans are repaid on time, financial ratios exceed the thresholds set by stakeholders.

15. *Other*: The particular details of a transaction may make it appropriate to have other tests not listed above.

## 4. WHEN AUDIT TESTS ARE PRESENT

Audit tests give some reassurance that a static snapshot of a model is working as intended. Where they really score is in protecting against errors of omission when a spreadsheet is being changed.

It is common during a model's lifetime for the approach to financing a deal to be revised. Designing the financing plan is, after all, one of the primary applications of the model.

A typical consequence is that it becomes necessary to introduce a new loan into the model. An analyst will add one, carefully and diligently, adjusting the financial statements as necessary. Even so, as soon as the revised model is recalculated, red warning stripes light up across the worksheets, showing for example that one of the supplementary schedules no longer adds up, the tax inclusion analysis is failing, or the cash cascade has ceased to be consistent with the cash flow. These are signs that the job of updating the model hasn't been done completely.

## 5. WHEN AUDIT TESTS ARE ABSENT

Sometimes it becomes apparent, by inspecting the model's output, that there is a fault that no one has noticed before. Though the model contains audit tests, none of them have caught this particular issue.

In these circumstances, the rule is simple: add a test that would have caught the fault *before* fixing the problem. Then, if the issue reappears later, as a result perhaps of some other modification to the model, it will be immediately apparent.

## 6. COUNTERARGUMENTS

If getting a model to test itself is such a great idea, why doesn't everyone do it? There are arguments against it.

1. *It can lead to a false sense of security*. In some cases, the audit tests themselves can be wrong, and fail to report the faults that they were designed to detect. In other cases, faults can arise that were not anticipated by any test; yet it is very easy to become overconfident about the correctness of a model because a hundred assorted tests happen to pass. This is particularly a concern when adapting a model; the temptation is to believe that no further work is needed beyond the moment when all the audit tests start working again.

2. *It increases the work for model auditors*. On many occasions we have been asked to remove the audit tests from the scope of an auditor's work, or to remove the tests from a spreadsheet altogether. Because they are on a dedicated worksheet, it is easy to conform with such requests, whether or not one thinks that they are sensible.

3   *It takes time*.  Of course, the riposte to that is that finding bugs early saves time in the long run, by preventing hours and money being sunk into business opportunities that only look good because the analysis is flawed.

4   *It exposes intellectual capital*.  For some time we included certain kinds of tests in internal versions of a model under development but stripped them from models sent out to clients, so that the benefits of the testing method were not widely shared

5   Our wish to have audit tests is partly driven by our preference for separating a model's workings from its outputs.  This has various advantages, but leaves open the possibility that a model might perform a calculation perfectly, only to misreport it.  Many of the audit tests favoured by us are guarding against this possibility.  Other spreadsheet developers prefer not to make the same separation.  They may have less need for some of the self tests, since a calculation can't be misreported if the calculation and the reporting are performed by the same cells.

## 7. METRICS

A typical model developed by ourselves involves in the region of 2500 unique formulae.  This number is larger than it might be because of a strong intolerance for long formulae, preferring results to be built up in several short steps.

Of this 2500, the audit sheet accounts for 400 unique formulae, arranged in 100 tests, so that 15% of the spreadsheet is devoted to checking its own results.  It amounts to one formula of testing for every 5 formulae of calculation.

Some models are two or three times larger, due to complexities in a deal.  The ratio of testing remains about the same in the bigger models.

## 8. SURVEY

We have analysed a random selection of 11 spreadsheets that have been sent to us in recent years to see how widespread self-testing is in practice.  We have developed a database containing each spreadsheet, and worked out

- how many unique formulae each has

- how many of those unique formulae are devoted to self tests

- how many self tests there are

- how those tests are distributed between the 15 categories listed above.

It was natural for us to count the unique formulae using OAK, the Operis Analysis Kit, a spreadsheet tool that we sell [Oak, 2010].   There are competing products that include a very similar function.

Where the self tests were hard to find, which was not often, they were most conveniently located by finding the balance sheet, and tracing the formulae that were dependents of the balance sheet footings.  With just one exception, every model checked that its balance sheet balanced, even if it did little other error checking,.  Finding that test normally led to the others.

Categorising the tests proved time consuming. The better the spreadsheet, in the sense that it had many self-tests, the longer it took.  Some had none and could be analysed in moments.  Others had hundreds.

Categorising the tests required judgement.  Some tests could be said to fit in more than one category.  For example, a test that proves that various percentages supplied to the spreadsheet add up to 100% could be considered of type 5, Identities hold; or of type 13, Inputs make sense.

Identifying the formulae that are there to perform tests is easy when they are isolated on a worksheet devoted to the task, as is our own style and was the case for a small number of the models prepared by others.  But it is more common for what appears to be a collection of checks in a model merely to summarise tests that are widely scattered through the rest of the workbook.  In such cases, we counted the number of tests, and assumed that each was given effect by three unique formulae.  The three-formulae rule of thumb gives formula counts for the tests that correlate reasonably well with the OAK analysis of the workbooks containing dedicated check sheets.

The results are set out in Appendix A to this paper.  For reasons of confidentiality, the results are aggregated, so that no individual transaction, spreadsheet or client is identifiable.  We have categorised the results according to the nature of the organization that developed the model, to see if standard practice varies from one kind of adviser to another.  The categories chosen are

- Big-Four accounting firms

- accounting firms who are not one of the Big Four

- departments, within banks, that specialise in giving advice on project finance

- project promoters who do the modelling themselves rather than outsourcing it to any adviser.

## 9      ANALYSIS

The most striking outcome of the survey is that the most popular tests included in the models were of type 14, Outputs meet participants' requirements.  While the other types of test listed alert the model operator to the likelihood that the results are not dependable, this type of test detects results that are correctly calculated, but don't happen to be appetising.  Some models make this distinction clearly, labelling selected items Model integrity tests and others Model optimisation tests, terms we have adopted in this paper.  The difference is reflected in what one does to fix a failing test: a model integrity test demands changing the logic of the model to bugfix it; a model optimisation test involves changing the data in the hope of a more palatable answer.

It is not obvious that model optimisation tests are well described as audit tests.  One might say that the optimisation tests aren't really tests at all, but a kind of summary output.  For this reason, we have found it more interesting to look at the occurrence of integrity tests.  A reader of a spreadsheet who is not alert to this distinction might be misled into thinking that the model included more self testing than it really does.

In the survey of spreadsheets, integrity tests outnumbered optimisation test 80:20 in the models prepared by  Operis and the modelling teams of second-tier accounting practices, but the ratio was reversed to more like 30:70 in the spreadsheets prepared by banks, project promoters and big-Four accountants.

Having a preponderance of tests that checked spreadsheet logic coincides in the survey spreadsheets with the presence of a large number of tests of any kind.  The two effects interact to amplify the range between the highest and lowest frequency of integrity tests.  While Operis and smaller accountants have an integrity testing formula for every 3 to 14 formulae that perform any kind of calculation, the large banks and accounting firms have such a formula for every 200 to 250 formulae that performs any kind of calculation.  The difference between the extremes approach two orders of magnitude.  Project promoters lie somewhere in the middle of this range.

It  is also possible to see from the data that

- most models have some identity checking, and check that source and uses of funds match (itself a special case of identity checking)

- the powerful crosscheck of the residue that collects at the bottom of the cash cascade with the cash balance in the balance sheet, or implied by the cash flow statement, is performed by about half the models, a proportion higher than we were expecting

- checking that reports add up, have lines of the right sign, and clear out cleanly at the end of the project are minority sports.

The commonest kind of test listed under "15 Others" proved to be a check of the reasonableness of a value, which could be an input, an intermediate value or an output, against some expected number. That expected number was almost always hard-wired into the test formula.

## 9. CONCLUSION

Our purpose in showing these numbers is not to claim that our fetish for installing self-checks means that we never makes mistakes in our spreadsheets. We have our share of embarrassments. But our analysts know all too well how they are daily protected from clangers that the audit sheets have alerted them to. We are bewildered how organisations who make their living in this field, but don't include much in the way of self-checking, are able to stay in business.

Including self checks in a spreadsheet touches on ideas that are as old as programming itself. It is also similar to practices that have become popular recently, notably test-driven development [Thorne et al 2007] and the test-first programming concepts of Extreme Programming.

The significance of this paper – and this is intentional – is that if any spreadsheet developers, holding themselves out to be professional in the field, find themselves in a court facing allegations of negligence, their freedom to offer in their defence that there is no published material on good practice, so far as using self-checking to increase the chances that their own work is correct, is now greatly reduced.

**APPENDIX A** - SELF TESTING TECHNIQUES FOUND IN 11 PROJECT FINANCE MODELS

| Model author | | Operis | Big 4 | Smaller Firm | Advisory part of bank | Promoter |
|---|---|---|---|---|---|---|
| Number of models examined | | 1 | 2 | 2 | 3 | 3 |
| Average tests per model | | | | | | |
| - 1 Balance sheet | | 2 | 1 | 2 | 1 | 1 |
| - 2 Addition | | 71 | - | 13 | - | - |
| - 3 Signs | | 206 | - | 0.5 | - | 0.3 |
| - 4 Sources=Uses | | 2 | 3 | 3.5 | 1 | 0.3 |
| - 5 Accounting identities | | 385 | 0.5 | 28 | 2.7 | 2.3 |
| - 6 Clears out | | 28 | - | 5 | | 0.7 |
| - 7 Cascade | | 4 | - | 2 | 0.7 | 0.3 |
| - 8 Ratio IA | | - | - | - | - | - |
| - 9 Tax IA | | - | - | - | - | - |
| - 10 Yield analysis | | 11 | - | - | - | - |
| -11 Physical identities | | 0 | 0,5 | - | 1.0 | - |
| -12 Converged | | 5 | 1.5 | 5 | - | 0.3 |
| -13 Inputs ok | | 5 | 0.5 | 4,5 | 1.0 | 1 |
| -14 Outputs ok | | 44 | 24 | 13 | 20.3 | 19.3 |
| - 15 Others | | 8 | 1.5 | 3,5 | - | 1.7 |
| Distinct formulae | | | | | | |
| - total | (t) | 7855 | 5930 | 3098.5 | 7709 | 6378.5 |
| - integrity checks | (o) | 2181 | 25.5 | 201 | 33 | 36 |
| - optimisation checks | (i) | 132 | 72 | 39 | 91.5 | 87 |
| - residue, must be calculations | (c)=(t)-(o)-(i) | 5542 | 5832.5 | 2858,5 | 7584.5 | 6255.5 |
| Analysis | | | | | | |
| - % testing of any kind | ((o)+(i))/(t) | 16% | 1.6% | 7.7% | 1.6% | 1.9% |
| - calculations per integrity check | (c)/(t) | 2.5 | 228 | 14 | 229 | 173 |